
\input amstex
\documentstyle{amsppt}

\hsize6.25truein
\vsize9.514truein

\parindent10pt
\output={\plainoutput}
\headline={\hfil}
\overfullrule0pt
\def\Hex#1{\ifcase#1 0\or1\or2\or3\or4\or5\or6\or7\or8\or9\or
 A\or B\or C\or D\or E\or F\fi}
\edef\Msa{\Hex\msafam}
\mathchardef\geqslant="3\Msa3E
\mathchardef\leqslant="3\Msa36
\mathchardef\square="0\Msa03

\def\gad#1{\global\advance#1 by1}
\newcount\Sno
\def\Subhead{\gad\Sno\Fno=0\subheading{\the\Sno}}
\newcount\Fno
\def\Tag#1{\expandafter\ifx\csname#1\endcsname\relax\gad\Fno
 \expandafter\xdef\csname#1\endcsname{\the\Sno.\the\Fno}\fi\tag{\[#1]}}
\def\[#1]{\csname#1\endcsname}
\def\({\allowbreak(}
\footline={\ifnum\pageno=1\hfil\else
  \ifodd\pageno\hfil\tenrm\folio\else\ninerm\folio\hfil\fi\fi}

\def\Lap{{\La{\kern-.15em}^{\sssize+}}}
\def\Lam{{\La{\kern-.15em}^{\sssize-}}}
\def\Lac{{\La{\kern-.15em}^\circ}}
\def\xiLa{{\xi_\La}}

\def\xiLam{{\xi_\Lam}}

\def\qqm{q-q^{-1}}

\def\Demo#1{\demo\nofrills{\smc#1.\usualspace}}
\def\Proof{\demo\nofrills{\it Proof\/.\usualspace}}
\def\ndots{\mathinner{\mkern1mu\raise1pt\hbox{.}\mkern2mu\raise4pt
\hbox{.}\mkern2mu\raise7pt\vbox{\kern7pt\hbox{.}}\mkern1mu}}
\def\ddd{\leaders\hbox to.6em{\hss.\hss}\hfil}
\def\gets{\vcenter{
 \hbox to 125pt{$\la_{n1}\ \ \la_{n2}\ddd\la_{nn}$}
 \hbox to 120pt{$\kern10pt\la_{n-1,1}\ \la_{n-1,2}\ddd\la_{n-1,n}$}
 \hbox to 89pt{$\kern22pt\ddots\hfil\ndots$}
 \hbox to 78pt{$\kern37pt\la_{21}\hfil\la_{22}$}
 \hbox{$\kern50pt\la_{11}$}}}

\def\C{{\Bbb C}}
\def\F{{\Bbb F}}
\def\pd{{\operatorname{pdet\thinspace}}}

\def\End{{\operatorname{End}}}
\def\enddemos{\ts$\square$\enddemo}
\def\gl{{\frak{gl}}}
\def\gn{{\gl_n}}
\def\id{{\operatorname{id}}}
\def\ot{{\otimes}}
\def\qd{{\operatorname{qdet\thinspace}}}
\def\Rc{{\check R}}
\def\SC{{\Cal S}}
\def\TC{{\Cal T}}
\def\Th{{\widehat{T}}}
\def\ts{\thinspace}
\def\Tt{{\widetilde{T}}}
\def\U{{\operatorname{U}}}
\def\Un{\U(\gn)}
\def\Uq{\U_{\!q}(\gn)}
\def\Y{{\operatorname{Y}}}
\def\Yn{\Y(\gn)}
\def\Yq{\Y\!_q(\gn)}
\def\Z{{\operatorname{Z}}}
\let\al\alpha
\let\be\beta
\let\ga\gamma
\let\de\delta
\let\eps\varepsilon
\let\la\lambda
\let\La\Lambda
\let\Om\Omega
\let\Up\Upsilon

\let\cl\centerline

\document

\font\bigbf=cmbx10 scaled 1440
\font\bigsl=cmsl10 scaled 1200
\line{\hfil\ninerm hep-th/93021**}
\cl{\bigsl RIMS \raise.16ex\hbox{--} 913}
\kern1.2truein
\cl{\bigbf Yangians and Gelfand-Zetlin bases}
\kern.4truein
\cl{Maxim NAZAROV\thinspace${}^\diamond$}
\kern.1truein
\cl{\sl Department of Mathematics, Moscow University}
\cl{\sl Moscow 119899, Russia\thinspace${}^\ast$}
\kern.15truein
\cl{and}
\kern.15truein
\cl{Vitaly TARASOV}
\kern.1truein
\cl{\sl Physics Department, Leningrad University}
\cl{\sl Leningrad 198904, Russia\thinspace${}^\ast$}
\kern1truein
\cl{\it To I.\ts M.\ts Gelfand on his 80\ts th birthday}
\kern1truein
\subheading{Abstact}
We establish a relationship between the modern theory of Yangians and the
classical construction of the Gelfand-Zetlin bases for the complex Lie algebra
$\gn$. Our approach allows us to produce the $q$-analogues of the
Gelfand-Zetlin formulae in a straightforward way.
\vskip 0pt plus 1filll
\line{\hrulefill}
\kern.05truein
\par\noindent
${}^\diamond$\thinspace
Japan Society for the Promotion of Science Postdoctoral Fellow
\par\noindent
${}^\ast$\thinspace
Current address:
Research Institute for Mathematical Sciences
\par\kern-.05cm\noindent\ \quad\quad\quad\quad\quad\quad\quad\quad\quad\quad
Kyoto University, Kyoto 606, Japan

\newpage

\noindent
Let $V$ be an irreducible finite-dimensional module over the complex Lie
algebra $\gn$. There is a canonical basis in the space of $V$ associated with
the chain of subalgebras
$\gl_1\subset\gl_2\subset\ldots\subset\gn\ts$.
It is called the Gelfand-Zetlin basis, and the action of $\gn$ on its vectors
was explicitly described in [GZ] for the first time. Since then several authors
provided alternative proofs of the original Gelfand-Zetlin formulae; see [Z2]
and references therein.
\par
Denote by $\Z(\gn)$  the centre of the universal enveloping algebra $\Un$.
The subalgebra in $\Un$ generated by
$\Z(\gl_1),\ts\Z(\gl_2),\ldots,\Z(\gn)$
is evidently commutative. The Gelfand-Zetlin basis in $V$ consists of the
eigenvectors of this subalgebra, and the corresponding eigenvalues are
pairwise distinct. This definition suggests that given the module $V$,
an explicit description of
$\Z(\gl_1),\ts\Z(\gl_2),\ldots,\Z(\gn)$
should be used to construct the Gelfand-Zetlin basis. It shall be done in the
present paper. Namely, for any vector $v$ of the Gelfand-Zetlin basis we point
out an element $b\in\Un$ such that $v=b\cdot\xi$ where $\xi\in V$ is the
highest weight vector (Section 2). Moreover, our construction implies
the Gelfand-Zetlin formulae (Section 3).
\par
We will employ the following description of $\Z(\gl_m)$, $m=1,\ldots,n$.
Let $e_{ij}$ be the standard generators of the algebra $\Un$.
Consider the sum over all permutations $g$ of $1,2,\ldots,m$
$$
\sum_g\ts\ts(-1)^{\ts\ell(g)}\prod_{i=1,\ldots,m}^\rightarrow
\bigl(\ts\de_{g(i),i}\ts(u-i+1)+e_{g(i),i}\ts\bigr)
$$
where $\ell(g)$ denotes the length of the permutation $g$. It is a polynomial
in $u$ and the coefficients of this polynomial generate $\Z(\gl_m)$ [Z1],
cf. [N].
\par
Our construction is based on the chain of subalgebras
$$
\Y(\gl_1)\subset\Y(\gl_2)\subset\ldots\subset\Yn
$$
where $\Y(\gl_m)$ denotes the Yangian [D1] of the Lie algebra $\gn$,
cf.~[C2]. In our construction we use the same generators of
the algebra $\Y(\gl_m)$ as were introduced in [D2], see also [T]. There exists
an algebra
homomorphism $\Y(\gl_m)\rightarrow\U(\gl_m)$. The generators of $\Z(\gl_m)$
mentioned above arise as the images of canonical central elements of
$\Y(\gl_m)$ with respect to that homomorhism (Section 1), cf.~[O].
\par
Our construction for $\Un$ also admits a natural generalization to the quantum
universal enveloping algebra $\Uq$, cf.~[C1]. This generalization produces
the same $q$-analogues of the Gelfand-Zetlin formualae as were
given in [J2]. To make the exposition clearer, we only formulate (Section 4)
the main statements for $\Uq$ and provide detailed proofs for $\Un$.

\Subhead
In this section we gather several known facts about the {\it Yangian} $\Yn$
of the complex Lie algebra $\gn$. This is an associative algebra generated
by the elements $T_{ij}^{(s)}$ where $i,j=1,\dots,n$ and $s=1,2,\dots$
subjected to the following relations. Introduce the formal Laurent series in
$u^{-1}$
$$
T_{ij}(u)=\de_{ij}\ts u+T_{ij}^{(1)}+
                           T_{ij}^{(2)}\ts u^{-1}+T_{ij}^{(3)}\ts u^{-2}+\ldots
$$
and form the matrix
$$
T(u)=[\ts T_{ij}(u)\ts]\ts_{i,j=1}^n.
$$
Let $P$ be the permutation map in $(\C^n)^{\ot2}\ts$. Consider the
{\it Yang $R$-matrix}, it is the
$\End((\C^n)^{\ot2})$-valued function
$$
R(u,v)=\id+\frac P{u-v}.
$$
Put $\Rc(u,v)=P\cdot R(u,v)$. Then the relations for
$T_{ij}^{(s)}$ can be written as
$$
\Rc(u,v)\cdot T(u)\ot T(v)=T(v)\ot T(u)\cdot\Rc(u,v).
\Tag{1.1}
$$
\par
Observe that the generators $T_{ij}^{(s)}$ with $i,j=1,\dots,m$ obey exactly
the
same relations as the corresponding generators of $\Y(\gl_m)$. Thus we have
the chain of subalgebras
$$
\Y(\gl_1)\subset\Y(\gl_2)\subset\ldots\subset\Yn.
$$
The relations (\[1.1]) also imply that for any $h\in\C$ the map
$$
T_{ij}(u)\mapsto T_{ij}(u+h)
$$
defines an automorhism of the algebra $\Yn$;
here the series in $(u+h)^{-1}$ should be re-expanded in $u^{-1}$.
\par
We will use the following definition.
Let $X(u)=[\ts X_{ij}(u)\ts]\ts_{i,j=1}^m$ be an arbitrary matrix whose entries
are formal Laurent series in $u^{-1}$ with coefficients from $\Yn$. Define the
{\it quantum determinant} of this matrix to be the sum over all permutations
$g$ of $1,2,\dots,m$
$$
\qd X(u)=\sum_g\ts(-1)^{\ell(g)}\cdot
X_{1g(1)}(u)\ts X_{2g(2)}(u-1)\dots X_{m,g(m)}(u-m+1);
$$
here $\ell(g)$ denotes the length of the permutation $g$. We will also denote
by $\pd X(u)$ the sum
$$
\sum_g\ts(-1)^{\ell(g)}\cdot
X_{1g(1)}(u-m+1)\ts X_{2g(2)}(u-m+2)\ldots X_{m,g(m)}(u).
$$
Consider the formal series
$$
A_m(u)=\qd[\ts T_{ij}(u)\ts]\ts_{i,j=1}^m;\quad m=1,\dots,n.
$$

\proclaim{Proposition 1.1}
a) The coefficients of $A_n(u)$ belong to the centre of the algebra
$\Yn$. b) All the coefficients of $A_1(u),\dots,A_n(u)$ pairwise commute.
\endproclaim

\Proof
The part a) is well known and its proof can be found for instance in [KS].
Since the generators $T_{ij}^{(s)}$ with $i,j=1,\dots,m$ obey the
same relations as the corresponding generators of $\Y(\gl_m)$,
we obtain from a) that
$$
[A_m(u),T_{ij}(v)]=0;\quad i,j=1,\ldots,m.
\Tag{1.2}
$$
The part b) follows directly from the above commutation relations
\enddemos
\noindent
It is convenient to assume $A_0(u)=1$.
Now we introduce the formal series with coefficients in $\Yn$ which
together with $A_1(u),\dots,A_n(u)$ play the main role in this paper. For any
$m=1,\dots,n-1$ denote by $B_m(u)$, $C_m(u)$, $D_m(u)$ respectively
the quantum determinants of the submatrices in $T(u)$ specified by the rows
$1,\dots,m$ and the columns $1,\dots,m-1,m+1$, by the rows $1,\dots,m-1,m+1$
and the columns $1,\dots,m$, by the same rows and the columns
$1,\dots,m-1,m+1$. For the first time so defined series had appeared in [D2].

\proclaim{Proposition 1.2}
The following commutation relations hold in $\Yn$:
$$
\align
[A_m(u),B_l(v)]&=0\quad\text{if}\quad l\neq m,
\Tag{1.3}
\\
[C_m(u),B_l(v)]&=0\quad\text{if}\quad l\neq m,
\Tag{1.4}
\\
[B_m(u),B_l(v)]&=0\quad\text{if}\quad|l-m|\neq1,
\Tag{1.5}
\endalign
$$
\vskip-1.2cm
$$
\gather
\ts\ts\ts\ts\ts
(u-v)\cdot[A_m(u),B_m(v)]=B_m(u)\ts A_m(v)-B_m(v)\ts A_m(u),
\Tag{1.6}
\\
\ts\ts\ts\ts\ts\ts
(u-v)\cdot[C_m(u),B_m(v)]=D_m(u)\ts A_m(v)-D_m(v)\ts A_m(u).
\Tag{1.7}
\endgather
$$
\endproclaim

\Proof
It follows from (\[1.1]) that the entries of any $m\times m$ submatrix in
$T(u)$
obey the same relations as the corresponding entries of the matrix
$[\ts T_{ij}(u)\ts ]\ts_{i,j=1}^m$. Therefore if we have two square submatrices
$X(u)$, $Y(u)$ in $T(u)$ and one of them contains the other, then due to
(\[1.2])
$$
[\qd X(u),\qd Y(v)]=0.
$$
This observation provides the relations (\[1.3]),\(\[1.4]),\(\[1.5]).
\par
It suffices to prove the relations (\[1.6]),\(\[1.7]) only for $m=n-1$.
Introduce the matrix
$$
\Th(u)=[\ts\Th_{ij}(u)\ts]\ts_{i,j=1}^n
$$
where $\Th_{ij}(u)$ is equal to $(-1)^{i-j}$ times the quantum determinant of
the matrix obtained from $T(u)$ by removing the row $j$ and the column $i$.
Then
$$
\gather
T(u)\ts\Th(u-1)=\qd T(u)\ts,
\Tag{1.8}\\
\noalign{\vskip2pt}
\Th^{\,t}(u)\,T^{\,t}(u-n+1)=\qd T(u)
\Tag{1.81}
\endgather
$$
where the superscript $t$ denotes the matrix transposition;
see [KS] for the proof of these equalities.
The matrix $T(u)$ is invertible as a formal Laurent series in $u^{-1}$; denote
by $\Tt(u)$ the inverse matrix. Then from (\[1.1]) we get the equality
$$
\Tt(u)\ot\Tt(v)\cdot\Rc(u,v)=\Rc(u,v)\cdot\Tt(v)\ot\Tt(u).
$$
The series $\qd T(u)$ is also invertible and commutes with each entry of the
matrix $\Th(u)$. Therefore from the last equality, (\[1.8]) and from
$$
\Rc(u+1,v+1)=\Rc(u,v)
$$
we obtain the matrix relation
$$
\Th(u)\ot\Th(v)\cdot\Rc(u,v)=\Rc(u,v)\cdot\Th(v)\ot\Th(u).
$$
\par
By the definition of the matrix $\Th(u)$ we have the equalities
$$
\align
A_{n-1}(u)=\Th_{nn}(u),
&\quad
D_{n-1}(u)=\Th_{n-1,n-1}(u),
\Tag{1.9}
\\
B_{n-1}(u)=-\ts\Th_{n-1,n}(u),
&\quad
C_{n-1}(u)=-\ts\Th_{n,n-1}(u).
\endalign
$$
The commutation relations (\[1.6]),\(\[1.7]) with $m=n-1$ are contained in the
above matrix relation
\enddemos
\noindent
We will keep using the matrices $\Th(u)$ and $\Tt(u)$ introduced in the proof
of Proposition 1.2.

\proclaim{Lemma 1.3}
 For any $m=1,\ldots,n-1$ we have the equality
$$
\qd[\ts T_{ij}(u-n+m)\ts]\ts_{i,j=1}^m=
\pd[\ts\Tt_{ij}(u)\ts]\ts_{i,j=m+1}^n\cdot\qd T(u).
$$
\endproclaim

\Proof
Let $w_1,w_2,\ldots,w_n$ be the standard basis in $\C^n$.
Let $I$ be the $n\times n$ matrix unit and  $J\in\End((\C^n)^{\ot n})$ be
the antisymmetrization map. Then
$$
T(u)\ot T(u-1)\ot\ldots\ot T(u-n+1)\cdot J=J\cdot\qd T(u)\ts;
$$
see [KS] for the proof of this equality. It implies that
$$
\align
I^{\ot(n-m)}&\ot
T(u-n+m)\ot\ldots\ot T(u-n+2)\ot T(u-n+1)\cdot J
\\
&=\Tt(u-n+m+1)\ot\ldots\ot\Tt(u-1)\ot\Tt(u)
\ot I^{\ot m}\cdot J\cdot\qd T(u).
\endalign
$$
Thus we have a matrix equality over the space $(\C^n)^{\ot n}$.
Taking its the diagonal entry corresponding to the vector
$$
w_{m+1}\ot\ldots\ot w_{n-1}\ot w_n\ot w_1\ot\ldots\ot e_{m-1}\ot w_m\ts,
$$
we get the equality claimed by Lemma 1.3
\enddemos

\proclaim{Proposition 1.4}
The following relation holds in $\Yn$:
$$
C_m(u)\ts B_m(u-1)=D_m(u)\ts A_m(u-1)-A_{m+1}(u)\ts A_{m-1}(u-1).
$$
\endproclaim

\Proof
It suffices to prove Proposition 1.4 only for $m=n-1$. Applying Lemma 1.3
to $m=n-2$ and using the equalities (\[1.8]),\(\[1.9]) along with
Proposition 1.1(a),
we get
$$
\gather
A_n(u)\ts A_{n-2}(u-1)=
\qd T(u)\cdot\qd[\ts T_{ij}(u-1)\ts]\ts_{i,j=1}^{n-2}=
\\
\qd T(u)\cdot\pd[\ts\Tt_{ij}(u+1)\ts]\ts_{i,j=n-1}^n\cdot\qd T(u+1)=
\\
\qd T(u)\cdot\bigl(\ts\Tt_{n-1,n-1}(u)\ts\Tt_{nn}(u+1)
-\Tt_{n-1,n}(u)\ts\Tt_{n-1,n}(u+1)\ts\bigr)\cdot\qd T(u+1)=
\\
\Th_{n-1,n-1}(u-1)\ts\Th_{nn}(u)-\Th_{n-1,n}(u-1)\ts\Th_{n,n-1}(u)
=D_{n-1}(u-1)\ts A_{n-1}(u)-B_{n-1}(u-1)\ts C_{n-1}(u).
\endgather
$$
Due to the relation (\[1.7]) the right hand side of the above equalities
coincides
with
$$
\quad\quad\quad\ts\ts
D_{n-1}(u)\ts A_{n-1}(u-1)-C_{n-1}(u)\ts B_{n-1}(u-1).
$$
Thus Proposition 1.4 for $m=n-1$ is proved
\enddemos

\Subhead
Let $e_{ij}$ be the standard generators of the universal enveloping algebra
$\Un$. The algebra $\Yn$ contains $\Un$ as a subalgebra: the maps
$e_{ji}\mapsto T_{ij}^{(1)}$ define the embedding. One can also
define a homomorphism $\Yn\to\Un$ by
$$
T_{ij}(u)\mapsto \de_{ij}\ts u+e_{ji}.
$$
Denote the images of the series $A_m(u)$, $B_m(u)$, $C_m(u)$ and $D_m(u)$ under
this homomorhism by $a_m(u)$, $b_m(u)$, $c_m(u)$ and $d_m(u)$ respectively.
These images are polynomial in $u$ and
$$
\gather
\ts\ts\ts\ts
a_m(u)=u^m+\left(
e_{11}+\ldots+e_{mm}-m(m-1)/2
\right)\ts u^{m-1}+\dots\ts,
\Tag{2.0}
\\
\ts\ts\ts\ts
b_m(u)=e_{m+1,m}\ts u^{m-1}+\dots,\quad c_m(u)=e_{m,m+1}\ts u^{m-1}+\dots\ts.
\endgather
$$
The above equalities show that the coefficients of the polynomials
$a_m(u)$, $b_m(u)$ and $c_m(u)$ generate the algebra $\Un$. We will
explicitly describe the action of these polynomials in each irreducible
finite-dimen\-si\-on\-al module of the Lie algebra $\gn$. We will use
Proposition 1.2 and Proposition 1.4 along with the following observation: if
$\frak{n}$ is the subalgebra in $\gl_m$ spanned by the elements $e_{ij},
\ts1\leqslant i<j\leqslant m$ then
by the definition of the quantum determinant
$$
\gather
a_m(u)\in\prod_{i=1}^m\left(u+e_{ii}-i+1\right)+\U(\gl_m)\ts\frak{n}\ts,
\Tag{2.1}
\\
c_m(u)\in\U(\gl_m)\ts\frak{n}\ts.
\Tag{2.2}
\endgather
$$
\par
Let $V$ be an irreducible finite dimensional $\gn$-module. Denote by $\xi$ its
highest weight vector:
$$
e_{ii}\cdot\xi=\la_i\ts\xi;\qquad e_{ij}\cdot\xi=0,\quad i<j.
$$
Then each difference $\la_i-\la_{i+1}$ is a non-negative integer.
 For any $h\in\C$ the map
$$
e_{ii}\mapsto e_{ii}+h;\qquad e_{ij}\mapsto e_{ij},\quad i\neq j
$$
define an automorphism of the algebra $\U(\gn)$. So we will assume that each
$\la_i$ is also an integer.
Denote by $\TC$ the set of all arrays $\La$ with integral entries of the form
$$
\gets
$$
where $\la_{ni}=\la_i$ and $\la_i\geqslant\la_{mi}$ for all $i$ and $m$.
The array $\La$ is called a {\it Gelfand-Zetlin scheme} if
$$
\la_{mi}\geqslant\la_{m-1,i}\geqslant\la_{m,i+1}
$$
for all possible $m$ and $i$. Denote by $\SC$ the subset in $\TC$
consisting of the Gelfand-Zetlin schemes.
\par
There is a canonical decomposition of the space $V$ into the direct sum of
one-dimensional subspaces associated with the chain of subalgebras
$$
\gl_1\subset\gl_2\subset\ldots\subset\gn.
$$
These subspaces are parametrized by the elements $\La\in\SC$. The subspace
$V_\La\subset V$ corresponding to $\La\in\SC$ is contained in an irreducible
$\gl_m$-submodule of the highest weight $(\la_{m1},\la_{m2},\ldots,\la_{mm})$
for each $m=n-1,n-2,\ldots,1$. These conditions define $V_\La$ uniquely [GZ].
Denote by $\Lac$ the array where $\la_{mi}=\la_i$ for any $m$; then
$\Lac\in\SC$ and $\xi\in V_{\Lac}$.
\par
 For any $\La\in\TC$ put
$$
\al_{m\La}(u)=\prod_{i=1}^m\left(u+\la_{mi}-i+1\right).
$$

\proclaim{Proposition 2.1}
The subspace $V_\La\subset V$ is an eigenspace of $a_m(u)$ with the eigenvalue
$\al_{m\La}(u)$.
\endproclaim

\Proof
The coefficients of the polynomial $a_m(u)$ belong to the center of the
algebra $\U(\gl_m)$ and act in an irreducible $\gl_m$-submodule of $V$ via
scalars. Applying $a_m(u)$ to the highest weight vector in this submodule and
using of (\[2.1]) we get Proposition 2.1 by the definition of the
subspace $V_\La$
\enddemos
\noindent
Endow the set of the pairs $(m,i)$ with the following relation of precedence:
$(m,i)\prec (l,j)$ if $i<j$ or $i=j$ and $m>l$. This relation corresponds to
reading
$\La\in\TC$ by diagonals from the left to the right, downwards in each
diagonal. Let $\nu_{mi}=i-\la_{mi}-1$, it is a root of the polynomial
$\al_{m\La}(u)$. Note that if $\La\in\SC$ then
$\nu_{m1}<\nu_{m2}<\ldots<\nu_{mm}$.
Put $\nu_i=i-\la_i-1$, then $\nu_{mi}\geqslant\nu_i\ts$. Consider the vector
in $V$
$$
\xiLa=\prod_{(l,j)}^\rightarrow
\Bigg(\ts\prod_{s=\nu_j}^{\nu_{lj}-1}\ts b_l(s)\Bigg)\cdot\xi\ts;
\Tag{2.3}
$$
here for each fixed $l$ the elements $b_l(s)\in\Un$ commute because of the
relation (\[1.5]).

\proclaim{Theorem 2.2}
 For any $\La\in\TC$ we have the equality
$$
a_m(u)\cdot\xiLa=\al_{m\La}(u)\ts\xiLa\ts.
$$
\endproclaim

\Proof
We will employ the induction on the number of the factors $b_l(s)$ in (\[2.3]).
If there is no factors then $\La=\Lac$ and $\xiLa=\xi\in V_{\Lac}$. In
particular, the required equality then holds by Proposition 2.1.
\par
Assume that $\La\neq\Lac\ts$. Let $(l,j)$ be the minimal pair such that
$\la_{lj}\neq\la_j$. Let $\Om$ be the array obtained from $\La$ by increasing
the $(l,j)$-entry by $1\ts$. Then $\Om\in\TC$ and
$$
\xiLa=b_l(\nu_{lj}-1)\cdot\xi_{\Om}\ts.
\Tag{2.4}
$$
If $l\neq m$ then $\al_{m\Om}(u)=\al_{m\La}(u)$. By the
relation (\[1.3]) and by the inductive assumption we get
$$
a_m(u)\cdot\xiLa=
a_m(u)\ts b_l(\nu_{lj}-1)\cdot\xi_{\Om}=
b_l(\nu_{lj}-1)\ts a_m(u)\cdot\xi_{\Om}
=b_l(\nu_{lj}-1)\cdot\al_{m\Om}(u)\ts
\xi_{\Om}=\al_{m\La}(u)\ts\xiLa\ts.
$$
\par
Now suppose that $l=m\ts$; then by the definition of $\Om$ we have
$$
\al_{m\Om}(u)=
\frac{u-\nu_{mj}+1}{u-\nu_{mj}}\ts\ts
\al_{m\La}(u).
$$
In particular, by the inductive assumption we then have
$$
a_m(\nu_{m j}-1)\cdot\xi_{\Om}=
\al_{m\Om}(\nu_{m j}-1)\ts\xi_{\Om}=0.
$$
Therefore by the relation (\[1.6]) and again by the inductive assumption we get
$$
\gather
a_m(u)\cdot\xiLa=
a_m(u)\ts b_m(\nu_{m j}-1)\cdot\xi_{\Om}
=\frac{u-\nu_{m j}}{u-\nu_{m j}+1}\ts\ts
b_m(\nu_{m j}-1)\ts a_m(u)\cdot\xi_{\Om}=
\\
\frac{u-\nu_{m j}}{u-\nu_{m j}+1}\ts\ts
b_m(\nu_{m j}-1)\cdot\al_{m\Om}(u)\ts\xi_{\Om}=
\al_{m\La}(u)\ts\xiLa\ts.
\endgather
$$
Thus Theorem 2.2 is proved for any $m$
\enddemos
\noindent
The subspaces $V_\La\subset V$ are separated by the corresponding
eigenvalues of $a_1(u),\ldots,a_{n-1}(u)$. Therefore by comparing Proposition
2.1 and Theorem 2.2 we get

\proclaim{Corollary 2.3}
 For any $\La\in\SC$ we have $\xiLa\in V_\La\ts$.
\endproclaim
\noindent
In the next section we will describe the action of the polynomials $b_m(u)$
and $c_m(u)$ on the vectors $\xiLa$ with $\La\in\SC$. Then we will prove
that all these vectors do not vanish.

\proclaim{Proposition 2.4}
If $\La\in\TC\setminus\SC$ then $\xi_{\La}=0$.
\endproclaim

\Proof
As well as in the proof of Theorem 2.2 we will employ the induction on the
number of the factors $b_l(s)$ in (\[2.3]). If there is no factors then
$\La=\Lac\in\SC$ and we have nothing to prove.
\par
Assume that $\La\neq\Lac\ts$. Let $(l,j)$ be the minimal pair such that
$\la_{lj}\neq\la_j$. Denote by $\Om$ the array obtained from increasing the
$(l,j)$-entry of $\La$ by $1\ts$, then $\Om\in\TC$ and we have the equality
(\[2.4]). If $\Om\notin\SC$ then $\xi_{\Om}=0$ by the inductive assumption, so
that $\xi_{\La}=0$ due to (\[2.4]).
\par
Now suppose that $\Om\in\SC$. We will prove that either $\La\in\SC$ or
$\xiLa=0$. By Theorem 2.2 the vector $\xiLa$ is an
eigenvector for the polynomials $a_1(u),\ldots,a_{n-1}(u)$. Their eigenvalues
separate the subspaces $V_\Up\subset V$ with $\Up\in\SC$. Therefore
$\xiLa\in V_\Up$ for some $\Up\in\SC$. Suppose that $\xiLa\neq0$, then
$$
a_{m\La}(u)=a_{m\Up}(u),\qquad m=1,\ldots,n-1.
$$
\par
Consider the roots $\nu_{mi}=i-\la_{mi}-1$ of the polynomial $a_{m\La}(u)$.
Since $\Om\in\SC$, we have the inequalities
$$
\nu_{m1}<\nu_{m2}<\ldots<\nu_{mm}\quad\text{if}\ \ m\neq l\,;\qquad
\nu_{l1}<\ldots<\nu_{lj}\leqslant\nu_{l,j+1}<\ldots<\nu_{ll}\ts.
$$
Therefore the array $\La$ can be uniquely restored from the collection of the
polynomials $\al_{1\La}(u),\ldots,\allowmathbreak\al_{n-1,\La}(u)$.
Thus $\La=\Up\in\SC$ and the Proposition 2.4 is proved
\enddemos

\Demo{Remark 2.5}
Let us form the matrix $E=[\ts-e_{ji}\ts]\ts_{i,j=1}^n\ts$. The coefficients
of the polynomial $a_n(u)=\qd(u-E)$ belong to the center of the
algebra $\Un$ and from (\[1.8]),\([1.81]) we obtain the matrix identities
$$
a_n(E)=0\,,\qquad a_n(E^{\,t\!}+n-1)=0\,.
\Tag{2.z}
$$
Now consider the matrix $S=[\ts S_{ij}\ts]\ts_{i,j=1}^n\ts$ where $S_{ij}$
denotes the image of the element $-e_{ji}$ in the module $V$.
Replacing the coefficients of the polynomial $a_n(u)$ by their eigenvalues in
the module $V$ and taking into account Proposition 2.1, we get the
{\it characteristic identities} [G] for the Lie algebra $\gn\ts$:
$$
\prod_{i=1}^n\left(S+\la_i-i+1\right)=0\,,\qquad
\prod_{i=1}^n\left(S^{\,t\!}+\la_i-i+n\right)=0\,.
$$
\enddemo

\Subhead
Let $\La\in\SC$ be fixed. The functions $b_m(u)\cdot\xiLa$ and
$c_m(u)\cdot\xiLa$ are polynomial in $u$ of the degree $m-1$. Since
$\nu_{m1}<\nu_{m2}<\ldots<\nu_{mm}\ts$, to describe these functions it suffices
to determine their values at $u=\nu_{mi}$ for each $i=1,\ldots,m$. This will be
done in the present section. Put
$$
\ga_{mi\La}=\ts
\prod_{j=1}^i\ts(\nu_{mi}-\nu_j)\ts
\prod_{j=1}^{i-1}\ts(\nu_{mi}-\nu_j-1)\times
\prod_{j=i+1}^{m+1}\ts(\nu_{m+1,j}-\nu_{mi})\ts
\prod_{j=i}^{m-1}\ts(\nu_{m-1,j}-\nu_{mi}+1)\ts.
$$
\par
Let the indices $m<n$ and $i\leqslant m$ be fixed. Denote by $\Lap$ the array
obtained from $\La$ by increasing the $(m,i)$-entry by $1$.

\proclaim{Theorem 3.1}
We have
$$
c_m(\nu_{mi})\cdot\xiLa=
\cases
\ga_{mi\La}\ts\xi_{\Lap}\quad&\text{if}\ \Lap\in\SC\ts;
\\
0\quad&\text{otherwise}.
\endcases
$$
\endproclaim

\Proof
If $\La=\Lac$ then $\Lap\notin\SC$. On the other hand then
$\xiLa=\xi$ while due to (\[2.2])
$$
c_m(\nu_{mi})\cdot\xi=0\ts.
$$
\par
Assume that $\La\neq\Lac\ts$. Consider the minimal pair $(l,j)$ such that
$\la_{lj}\neq\la_j$. As well as in the proof of Theorem 2.2 let $\Om$ be
the array obtained from $\La$ by increasing the $(l,j)$-entry by
$1\ts$. Then $\Om\in\SC$ and we have the equality (\[2.4]).
We will prove first that
$$
c_m(\nu_{mi})\ts b_l(\nu_{lj}-1)\cdot\xi_{\Om}\ts=
b_l(\nu_{lj}-1)\ts c_m(\nu_{mi})\cdot\xi_{\Om}
\Tag{3.1}
$$
for $\ts(l,j)\neq(m,i)\ts$.
If $l\neq m$ then we obtain (\[3.1]) directly from the relation (\[1.4]).
If $l=m$ but $j\neq i$ then $\nu_{mj}-1\neq\nu_{mi}$ since
$\Om\in\SC$. Due to Theorem 2.2 we then also have the equalities
$$
a_m(\nu_{mj}-1)\cdot\xi_{\Om}=\al_{m\Om}(\nu_{mj}-1)\ts\xi_{\Om}=
a_m(\nu_{mi})\cdot\xi_{\Om}=\al_{m\Om}(\nu_{mi})\ts\xi_{\Om}=0\ts.
$$
Therefore by the relation (\[1.7]) we again obtain that
$$
c_m(\nu_{mi})\ts b_m(\nu_{mj}-1)\cdot\xi_{\Om}\ts=
b_m(\nu_{mj}-1)\ts c_m(\nu_{mi})\cdot\xi_{\Om}\ts.
$$
\par
If $\la_{mi}=\la_i$ then $\Lap\notin\SC$. On the other
hand, applying the equality (\[3.1]) repeatedly we then get
$$
c_m(\nu_{mi})\cdot\xiLa=c_m(\nu_{mi})\ts\prod_{(l,j)}^\rightarrow
\Bigg(\ts\prod_{s=\nu_j}^{\nu_{lj}-1}\ts b_l(s)\Bigg)\cdot\xi
=\prod_{(l,j)}^\rightarrow\Bigg(
\ts\prod_{s=\nu_j}^{\nu_{lj}-1}\ts b_l(s)\Bigg)
c_m(\nu_{mi})\cdot\xi=0,
$$
as we have claimed. Now we assume that $\la_{mi}<\la_i$.
\par
Consider the array $\Up$ obtained from $\La$ by changing each entry
corresponding to $(l,j)\prec(m,i)$ for $\la_j\ts$, and by increasing
the $(m,i)$-entry by $1$. Then $\Up\in\TC$ and due to Theorem 2.2 we have
$$
a_m(\nu_{mi}-1)\cdot\xi_{\Up}=\al_{m\Up}(\nu_{mi}-1)\ts\xi_{\Up}=0\ts.
$$
We then also have
$$
\xiLa=p\ts b_m(\nu_{mi}-1)\cdot\xi_\Up\ts
$$
where
$$
p=\prod_{(l,j)\prec(m,i)}^\rightarrow
\Bigg(\ts\prod_{s=\nu_j}^{\nu_{lj}-1}\ts b_l(s)\Bigg).
$$
Therefore applying the equality (\[3.1]) repeatedly and using Proposition 1.4
we get
$$
\gather
c_m(\nu_{mi})\cdot\xiLa=p\ts c_m(\nu_{mi})\ts b_m(\nu_{mi}-1)\cdot\xi_\Up
=-p\ts a_{m+1}(\nu_{mi})\ts a_{m-1}(\nu_{mi}-1)\cdot\xi_\Up=\\
-\al_{m+1,\Up}(\nu_{mi})\ts\al_{m-1,\Up}(\nu_{mi}-1)\ts p\cdot\xi_\Up
=-\al_{m+1,\Up}(\nu_{mi})\ts\al_{m-1,\Up}(\nu_{mi}-1)\ts\xi_{\Lap}\ts
=\ga_{mi\La}\ts\xi_{\Lap}\ts.
\endgather
$$
The last equality proves Theorem 3.1 when $\Lap\in\SC$. If
$\Lap\notin\SC$ then
$$
c_m(\nu_{mi})\cdot\xiLa=0
$$
by the same equality and by Proposition 2.4
\enddemos
\noindent
\Demo{Remark 3.2}
If $\Lap\in\SC$ then $\ga_{mi\La}>0$. Indeed, if $\La\in\SC$ then we have the
inequalities
$$
\nu_{mi}<\nu_{m+1,i+1}<\nu_{m+1,i+2}<\ldots<\nu_{m+1,m+1}\ts;\qquad
\nu_{mi}\leqslant\nu_{m-1,i}<\nu_{m-1,i+1}<\ldots<\nu_{m-1,m-1}\ts.
$$
If $\Lap\in\SC$ then we also have
$$
\nu_{mi}-1\geqslant\nu_i>\nu_{i-1}>\ldots>\nu_1\ts.
$$
Thus all the factors in the product $\ga_{mi\La}$ are positive.
\enddemo

\proclaim{Proposition 3.3}
If $\La\in\SC$ then $\xiLa\neq0$.
\endproclaim

\Proof
As well as in the proofs of Theorem 2.2 and Proposition 2.4 we will employ the
induction on the number of the factors $b_l(s)$ in (\[2.3]).
If there is no factors then $\La=\Lac$ and $\xiLa=\xi\neq0$.
\par
Assume that $\La\neq\Lac\ts$. Let $(l,j)$ be the minimal pair such that
$\la_{lj}\neq\la_j$. Let $\Om$ be the array obtained from $\La$ by increasing
the $(l,j)$-entry by $1\ts$. Since $\La\in\SC$, we also have $\Om\in\SC$. Then
$\xi_\Om\neq0$ by the inductive assumption. On the other hand, by Theorem 3.1
we then have
$$
c_l(\nu_{lj})\cdot\xiLa=\ga_{lj\La}\ts\xi_\Om
$$
where $\ga_{lj\La}\neq0$ due to Remark 3.2. Therefore $\xiLa\neq0$ either
\enddemos
\noindent
Now consider the array $\Lam$ obtained from $\La$ by decreasing the
$(m,i)$-entry of $\La$ by $1$; then $\Lam\in\TC$. Since
$$
\nu_{mi}\geqslant\nu_i>\nu_{i-1}>\ldots>\nu_1\ts,
\Tag{3.2}
$$
one can define
$$
\be_{mi\La}=
\prod_{j=1}^i\ts\frac
{\nu_{mi}-\nu_{m+1,j}+1}{\nu_{mi}-\nu_j+1}\ts\ts
\prod_{j=1}^{i-1}\ts\frac
{\nu_{mi}-\nu_{m-1,j}}{\nu_{mi}-\nu_j}
$$

\proclaim{Theorem 3.4}
We have
$$
b_m(\nu_{mi})\cdot\xiLa=
\cases
\be_{mi\La}\ts\xi_{\Lam}\quad&\text{if}\ \ \Lam\in\SC\ts;
\\
0\quad&\text{otherwise}.
\endcases
$$
\endproclaim

\Proof
We will use again some of the arguments which appeared in the proofs of Theorem
2.2 and Proposition 2.4. Consider the vector $b_m(\nu_{mi})\cdot\xiLa\in V$.
It is an eigenvector of the element $a_l(u)$ with the eigenvalue
$\al_{l\Lam}(u)$
for any $l$. Indeed, if $l\neq m$ then $\al_{l\Lam}(u)=\al_{l\La}(u)$. On the
other hand, by the relation (\[1.3]) and by Theorem 2.2 we then get
$$
a_l(u)\ts b_m(\nu_{mi})\cdot\xi_{\La}=
b_m(\nu_{mi})\ts a_l(u)\cdot\xi_{\La}=
\al_{l\La}(u)\ts b_m(\nu_{mi})\cdot\xiLa\ts.
$$
Now suppose that $l=m\ts$; then by the definition of $\Lam$ we have
$$
\al_{m\Lam}(u)=
\frac{u-\nu_{mi}-1}{u-\nu_{mi}}\ts\ts
\al_{m\La}(u).
$$
Since $a_m(\nu_{mi})\cdot\xi_{\La}=0$ due to Theorem 2.2, by the relation
(\[1.6]) and again by Theorem 2.2 we get
$$
a_m(u)\ts b_m(\nu_{mi})\cdot\xi_{\La}=
\frac{u-\nu_{mi}-1}{u-\nu_{mi}}\ts\ts
b_m(\nu_{mi})\ts a_m(u)\cdot\xi_{\La}
=\al_{m\Lam}(u)\ts b_m(\nu_{mi})\cdot\xiLa\ts.
$$
\par
The subspaces $V_\Om\subset V$ with $\Om\in\SC$ are separated by the
eigenvalues of $a_1(u),\ldots,a_{n-1}(u)$. Therefore
$b_m(\nu_{mi})\cdot\xi_{\La}\in V_\Om$ for some $\Om\in\SC$. Since
$\La\in\SC$, the array $\Lam$ can be uniquely restored from the collection of
the polynomials $\al_{1\Lam}(u),\ldots,\al_{n-1,\Lam}(u)$. Thus if
$\Lam\notin\SC$ then
$$
b_m(\nu_{mi})\cdot\xi_{\La}=0
$$
as we have claimed.
\par
Assume that $\Lam\in\SC$. Then the above consideration implies that
$$
b_m(\nu_{mi})\cdot\xi_{\La}\in V_\Lam\ts.
$$
Therefore by Corollary 2.3 and Proposition 3.3 we obtain that
$$
b_m(\nu_{mi})\cdot\xi_{\La}=\be\ts\xiLam
\Tag{3.3}
$$
for some $\be\in\C$.
We will prove that $\be=\be_{mi\La}$.
\par
Let us compare the action of the element
$c_m(\nu_{mi}+1)$ on the both sides of the equality (\[3.3]). By Theorem 3.1 we
have
$$
c_m(\nu_{mi}+1)\cdot\be\ts\xiLam=\be\ts\ga_{mi\Lam}\ts\xiLa
$$
where $\ga_{mi\Lam}\neq0$ due to Remark 3.2. On the other hand, applying
Proposition 1.4 and using the equality
$$
a_m(\nu_{mi})\cdot\xiLa=\al_{m\La}(\nu_{mi})\ts\xiLa=0,
$$
we get
$$
c_m(\nu_{mi}+1)\ts b_m(\nu_{mi})\cdot\xi_{\La}=
-a_{m+1}(\nu_{mi}+1)\ts a_{m-1}(\nu_{mi})\cdot\xiLa
=-\al_{m+1,\La}(\nu_{mi}+1)\ts\al_{m-1,\La}(\nu_{mi})\ts\xiLa\ts.
$$
Since $\xiLa\neq0$ by Proposition 3.3, we finally obtain that
$$
\be=-\al_{m+1,\La}(\nu_{mi}+1)\ts\al_{m-1,\La}(\nu_{mi})\ts
\ga_{mi\Lam}^{-1}=\be_{mi\La}\ts.
$$
Thus we have proved Theorem 3.4
\enddemos

\Demo{Remark 3.5}
If $\La\in\SC$ then $\be_{mi\La}>0$ for any indices $m$ and $i$. Indeed, then
$$
\nu_{mi}\geqslant\nu_{m+1,i}>\nu_{m+1,i-1}>\ldots>\nu_{m+1,1}\ts;\qquad
\nu_{mi}>\nu_{m-1,i-1}>\nu_{m-1,i-2}>\ldots>\nu_{m-1,1}\ts.
$$
These inequalities along with (\[3.2]) show that all the factors in the product
$\be_{mi\La}$ are positive.
\enddemo
\noindent
Theorems 2.2,\ts3.1,\ts3.4 and Proposition 3.3 along with Corollary 2.3
completely describe the action of the Lie algebra $\gn$ in the module $V$.
In particular, they provide explicit formulae for the action of generators
$e_{mm}$, $e_{m,m+1}$ and $e_{m+1,m}$ on the vectors $\xiLa$ with $\La\in\SC$.
The first equality in (\[2.0]) and Theorem 2.2 imply that
\vadjust{\kern1pt}
$$
e_{mm}\cdot\xiLa=\Bigg(\sum_{i=1}^m\la_{mi}-\sum_{i=1}^{m-1}\la_{m-1,i}\Bigg)
\ts\xiLa\ts.
\Tag{3.z}
$$
Put
$$
\tau_{mi\La}=\prod^m_{{\ssize j=1}\atop{\ssize j\ne i}}
(\nu_{mi}-\nu_{mj})^{-1}\ts.
$$
\vskip1pt\noindent
Then using the Lagrange interpolation formula we obtain that
$$
e_{m,m+1}\cdot\xiLa=\sum_{\Lap}\ts\ga_{mi\La}\ts\tau_{mi\La}\ts\xi_{\Lap}\,,
\qquad
e_{m+1,m}\cdot\xiLa=\sum_{\Lam}\ts\be_{mi\La}\ts\tau_{mi\La}\ts\xi_{\Lam}
\Tag{3.0}
$$
\vskip1pt\noindent
where $\Lap$ and $\Lam$ are Gelfand-Zetlin schemes obtained from $\La$
by increasing and decreasing the $(m,i)$-entry by $1$ respectively.
\par
The equalities (\[3.0]) are not the Gelfand-Zetlin formulae in their canonical
form [GZ]. To obtain the latter, one should employ vectors which differ from
$\xiLa$ by certain scalar factors. Namely, one should multiply $\xiLa$ by
the factor
\vadjust{\kern1pt}
$$
\prod_{(l,j)}\kern6pt\Bigg\{\,\prod_{k=1}^{j-1}\,(\nu_{\,lj}-\nu_{\,lk})
\cdot\kern-2pt
\prod_{s=\nu_j}^{\nu_{lj}-1}\,\Bigg(
\prod_{k=1}^j\,(s+1-\nu_k)\,\prod_{k=1}^{j-1}(s-\nu_k)\,
\prod_{k=j+1}^{l+1}\,(\nu_{\,l+1,k}-s-1)\,
\prod_{k=j}^{l-1}\,(\nu_{\,l-1,k}-s)\Bigg)\Bigg\}^{-1/2}\kern-17pt.
$$
Since $\La\in\SC$ and $\nu_j\leqslant s<\nu_{\,lj}$, all factors in the above
products are positive.

\Subhead
The construction given above admits a natural generalization to the case of
the quantum universal enveloping algebra $\Uq$. We will point out here only the
main statements. The proofs are quite similar to those in the case of $\Un$
and will be omitted. Some of them are contained in [T].
\par
Let us introduce the {\it quantum Yangian} $\Yq$, cf.~[C1]. This is an
associative algebra over the field $\F=\Bbb Q(q)$, generated by the
elements $T_{ij}^{(s)}$ where $i,j=1,\dots,n$ and
$s=0,1,\dots$ such that $T_{ii}^{(0)}$ are invertible and $T_{ij}^{(0)}=0$ for
$i>j$. These elements are subjected to the following relations.
Introduce the formal Laurent series in $x^{-1}$
$$
T_{ij}(x)=T_{ij}^{(0)}\ts x+T_{ij}^{(1)}+
T_{ij}^{(2)}\ts x^{-1}+T_{ij}^{(3)}\ts x^{-2}+\ldots
$$
and form the matrix
$$
T(x)=[\ts T_{ij}(x)\ts]\ts_{i,j=1}^n.
$$
Let $w_1,\ldots,w_n$ be the standard basis in $\F^n$.
Consider the {\it Cherednik $R$-matrix}, it is the
$\End((\F^n)^{\ot2})$-valued function $R_q(x,y)$ such that
$$
R_q(x,y)\cdot w_i\ot w_j=
\cases
(xq-yq^{-1})\ts w_i\ot w_i,\quad&i=j\ts;
\\
(x-y)\ts w_i\ot w_j+x(\qqm)\ts w_j\ot w_i,\quad&i>j\ts;
\\
(x-y)\ts w_i\ot w_j+y(\qqm)\ts w_j\ot w_i,\quad&i<j\ts.
\endcases
$$
Let $P$ be the permutation map in $(\F^n)^{\ot2}\ts$;
put $\Rc(x,y)=P\cdot R(x,y)$. Then the relations for $T_{ij}^{(s)}$ are of the
same form as (\[1.1]) above:
$$
\Rc(x,y)\cdot T(x)\ot T(y)=T(y)\ot T(x)\cdot\Rc(x,y).
\Tag{4.a}
$$
\par
The generators $T_{ij}^{(s)}$ with
$i,j=1,\dots,m$ obey exactly the same relations as the corresponding
generators of $\Y_q(\gl_m)$. Thus we have the chain of subalgebras
$$
\Y\!_q(\gl_1)\subset\Y\!_q(\gl_2)\subset\ldots\subset\Yq.
$$
The relations (\[4.a]) also imply that for any $h\in\F\setminus\{0\}$ the map
$$
T_{ij}(x)\mapsto T_{ij}(xh)
$$
defines an automorhism of the algebra $\Yq$.
\par
Let $X(x)=[\ts X_{ij}(x)\ts]\ts_{i,j=1}^m$ be an arbitrary matrix whose entries
are formal Laurent series in $x^{-1}$ with coefficients from $\Yq$.
Define the {\it quantum determinant} of this matrix to be the sum over all
permutations $g$ of $1,2,\dots,m$
$$
\qd X(x)=\sum_g(-q)^{-\ell(g)}\!\cdot\!
X_{1g(1)}(x) X_{2g(2)}(xq^{-2})\dots X_{m,g(m)}(xq^{2-2m}).
$$
We will also denote by $\pd X(x)$ the sum
$$
\sum_g(-q)^{\ell(g)}\!\cdot\!
X_{1g(1)}(xq^{2-2m})\ts X_{2g(2)}(xq^{4-2m})\ldots X_{m,g(m)}(x).
$$
\par
Define the formal series $A_m(x)$, $B_m(x)$, $C_m(x)$ and $D_m(x)$ by the
matrix $T(x)$ in the same way as the formal series $A_m(u)$, $B_m(u)$, $C_m(u)$
and $D_m(u)$ were defined by the matrix $T(u)$ in Section 1.

\proclaim{Proposition 4.1}
a) The coefficients of $A_n(x)$ belong to the centre of the algebra
$\Yq$. b) All the coefficients of $A_1(x),\dots,A_n(x)$ pairwise commute.
\endproclaim
\noindent
Define the $q$-commutator $[X,Y]_q=XY-qY\!X$ as usual. Then we have

\proclaim{Proposition 4.2}
The following commutation relations hold in $\Yq$:
$$
\align
[A_m(x),B_l(y)]&=0\quad\text{if}\quad l\neq m,
\\
[C_m(x),B_l(y)]&=0\quad\text{if}\quad l\neq m,
\\
[B_m(x),B_l(y)]&=0\quad\text{if}\quad|l-m|\neq1,
\\
\frac{x-y\ }{\ \qqm}\ts[A_m(x),B_m(y)]_q&=
y\ts B_m(x)\ts A_m(y)-x\ts B_m(y)\ts A_m(x),
\\
\frac{x-y\ }{\ \qqm}\ts[C_m(x),B_m(y)]&=
y\ts(D_m(x)\ts A_m(y)-D_m(y)\ts A_m(x)).
\endalign
$$
\endproclaim
\noindent
The matrix $T(x)$ is invertible as a formal Laurent series in $x^{-1}$; denote
by $\Tt(x)$ the inverse matrix.

\proclaim{Lemma 4.3}
 For any $m=1,\ldots,n-1$ we have the equality
$$
\qd[\ts T_{ij}(x\kern.2pt q^{2(m-n)})\ts]\ts_{i,j=1}^m=
\pd[\ts\Tt_{ij}(x)\ts]\ts_{i,j=m+1}^n\cdot\qd T(x).
$$
\endproclaim

\proclaim{Proposition 4.4}
The following relation holds in $\Yq$:
$$
q\,C_m(x\kern.2pt q^2)\ts B_m(x)=D_m(x\kern.2pt q^2)\ts A_m(x)-
A_{m+1}(x\kern.2pt q^2)\ts A_{m-1}(x).
$$
\endproclaim
\noindent
By definition, the {\it quantum universal enveloping algebra} $\Uq$ is an
associative algebra over $\F$ generated by the elements $t_i$, $t_i^{-1}$ with
$i=1,\ldots,n$ and $e_i$, $f_i$ with $i=1,\ldots,n-1$. These elements are
subjected to the following relations [J1]:
$$
\gather
t_it_i^{-1}=t_i^{-1}t_i=1\,,\qquad[t_i,t_j]=0\,,\\
t_ie_j=e_jt_iq^{\de_{ij}-\de_{i,j+1}}\,,\qquad
t_if_j=f_jt_iq^{\de_{i,j+1}-\de_{ij}}\,,\\
\noalign{\vskip2pt}
[e_i,f_j]={t_it_{i+1}^{-1}-t_{i+1}t_i^{-1}\over\qqm}\ts\de_{ij}\,,\\
\noalign{\vskip2pt}
[e_i,e_j]=[f_i,f_j]=0\quad\text{if}\quad |i-j|>1\,,\\
\noalign{\vskip4pt}
[e_i,[e_{i\pm1},e_i]_q]_q=[f_i,[f_{i\pm1},f_i]_q]_q=0\,.
\endgather
$$
Introduce the $q$-analogues of the root vectors in $\gn$ by induction:
$$
\alignat3
e_{i,i+1}&=e_i\,,\kern3em&e_{ij}&=[e_{ik},e_{kj}]_q\qquad&&i<k<j\,,\\
\noalign{\vskip1pt}
e_{i+1,i}&=f_i\,,\kern3em&e_{ij}&=[e_{ik},e_{kj}]_{q^{-1}}\qquad&&i>k>j\,.
\endalignat
$$
One can define a homomorphism $\Yq\to\Uq$ as follows [J1]:
$$
T_{ii}(x)\mapsto{x\,t_i-t_i^{-1}\over\ \qqm}\,,\kern3em
T_{ij}(x)\mapsto x\ts t_ie_{ji}\ts,\qquad i<j\ts;\kern3em
T_{ij}(x)\mapsto e_{ji}\ts t_j^{-1}\ts,\qquad i>j\ts.
\Tag{Hom}
$$
Denote the images of the series $A_m(x)$, $B_m(x)$, $C_m(x)$ and $D_m(x)$ under
this homomorhism by $a_m(x)$, $b_m(x)$, $c_m(x)$ and $d_m(x)$ respectively.
These images are polynomial in $x$ and
$$
\aligned
a_m(x)&=\bigl(x^m\ts q^{m(1-m)}\ts\ts t_1\ldots t_m+\ldots+
(-1)^m\ts t_1^{-1}\!\!\ldots t_m^{-1}\bigr)\cdot\left(\qqm\right)^{-m}\!,
\\
b_m(x)&=\bigl(x^m\ts q^{m(1-m)}\ts\ts t_1\ldots t_m\ts
f_m+\ldots+x\ts b\bigr)\cdot\left(\qqm\right)^{1-m}\!,
\\
c_m(x)&=\bigl(x^{m-1}\ts c+\ldots+(-1)^m\ts e_m\ts t_1^{-1}\!\!\ldots
t_m^{-1}\bigr)\cdot\left(\qqm\right)^{1-m}
\endaligned
$$
for some $b,\,c\in\Uq$. The above equalities show that the coefficients of
the polynomials $a_m(x)$, $b_m(x)$ and $c_m(x)$ generate the algebra $\Uq$.
\par
Let us recall several known facts about finite-dimensional $\Uq$-modules
[J1,L,R].
It is known that any such module is completely reducible and all the
irreducible modules are uniquely characterized by their highest weights.
Let $V$ be an irreducible finite-dimensional $\Uq$-module of the highest weight
$(\kappa_1,\kappa_2,\ldots,\kappa_n)$.
Denote by $\xi$ the highest weight vector:
$$
t_i\cdot\xi=\kappa_{i\,}\xi\,;\qquad e_{ij}\cdot\xi=0,\quad i<j.
$$
Then $\kappa_i=\eps_{i\,}q^{\la_i}$ where $\eps_i=\pm1$, $\la_i\in\Bbb Z$ and
$\la_i\geqslant\la_{i+1}$. The maps
$$
t_i\mapsto\eps_i\ts t_i\ts,\qquad
e_i\mapsto\eps_i\ts e_i\ts,\qquad
f_i\mapsto\eps_{i+1}\ts f_i
$$
define an automorphism of $\Uq$. So we will assume that each $\eps_i=1\,$.
\par
There is a canonical decomposition of the space $V$ into the direct sum of
one-dimensional subspaces associated with the chain of subalgebras
$$
\U_{\!q}(\gl_1)\subset\U_{\!q}(\gl_2)\subset\ldots\subset\Uq.
$$
These subspaces are parametrized by the Gelfand-Zetlin schemes $\,\La$.
The subspace $V_\La\subset V$ corresponding to $\La\in\SC$ is contained in
an irreducible $\U_q(\gl_m)$-submodule of the highest weight
$(q^{\la_{m1}},q^{\la_{m2}},\ldots,q^{\la_{mm}})$
for each $m=n-1,n-2,\ldots,1$. These conditions define $V_\La$ uniquely.
\par
Let again $\nu_{mi}=i-\la_{mi}-1$. For any $\La\in\TC$ put
$$
\al_{m\La}(x)=\prod_{i=1}^m\,
{x\,q^{-\nu_{mi}}-q^{\nu_{mi}}\over\qqm}.
$$

\proclaim{Proposition 4.5}
The subspace $V_\La\subset V$ is an eigenspace of $a_m(x)$ with the eigenvalue
$q^{m(1-m)/2\,}\al_{m\La}(x)$.
\endproclaim
\noindent
 For any $\La\in\TC$ define the vector $\xiLa$ in a way similar (\[2.3]):
$$
\xiLa=\prod_{(l,j)}^\rightarrow
\Bigg(\ts\prod_{s=\nu_j}^{\nu_{lj}-1}\ts
q^{l(l-1)/2-sl+1}\,b_l(q^{2s})\Bigg)\cdot\xi\ts,
\Tag{4.b}
$$
here for each fixed $l$ the elements $b_l(q^{2s})\in\Uq$ commute because
of Proposition 4.2\,.

\proclaim{Theorem 4.6}
a) For any $\La\in\TC$ we have the equality
$$
\quad a_m(x)\cdot\xiLa=q^{m(1-m)/2\,}\al_{m\La}(x)\ts\xiLa\ts.
$$
b) For any $\La\in\SC$ we have $\,\xiLa\in V_\La\ts$.\newline
c) If $\La\in\TC\setminus\SC$ then $\,\xi_{\La}=0$.
\endproclaim
\noindent
 For any $k\in\Bbb Z$ put $\dsize[k]={q^k-q^{-k}\over\qqm}$ and define
$$
\align
\be_{mi\La}&=\prod_{j=1}^i\ts\frac
{[\nu_{mi}-\nu_{m+1,j}+1]}{[\nu_{mi}-\nu_j+1]}\ts\ts
\prod_{j=1}^{i-1}\ts\frac{[\nu_{mi}-\nu_{m-1,j}]}{[\nu_{mi}-\nu_j]}\,,\\
\ga_{mi\La}&=\ts\prod_{j=1}^i\ts[\nu_{mi}-\nu_j]\ts
\prod_{j=1}^{i-1}\ts[\nu_{mi}-\nu_j-1]\,
\prod_{j=i+1}^{m+1}\ts[\nu_{m+1,j}-\nu_{mi}]\ts
\prod_{j=i}^{m-1}\ts[\nu_{m-1,j}-\nu_{mi}+1]\ts\,,\\
\tau_{mi\La}&=\prod^m_{{\ssize j=1}\atop{\ssize j\ne i}}
 [\nu_{mi}-\nu_{mj}]^{-1}\,.
\endalign
$$
Let the indices $m<n$ and $i\leqslant m$ be fixed. Denote by $\Lap$
and $\Lam$ the arrays obtained from $\La$ by increasing and decreasing
the $(m,i)$-entry by 1 respectively.

\proclaim{Theorem 4.7}
a) We have
\vadjust{\kern-12pt}
$$
c_m(q^{2\nu_{mi}})\cdot\xiLa=
\cases
q^{m(1-m)/2+m\ts\nu_{mi}}\,
\ga_{mi\La}\ts\xi_{\Lap}\quad&\text{if}\ \Lap\in\SC\ts;
\\
0\quad&\text{otherwise}.
\endcases
$$
b) If $\La\in\SC$ then $\xiLa\neq0$.\newline
c) We have
$$
b_m(q^{2\nu_{mi}})\cdot\xiLa=
\cases
q^{m(1-m)/2+m\ts\nu_{mi}-1}\,
\be_{mi\La}\ts\xi_{\Lam}\quad&\text{if}\ \ \Lam\in\SC\ts;
\\
0\quad&\text{otherwise}.
\endcases
$$
\endproclaim
\noindent
Theorems 4.6,\ts4.7 completely describe the action of the algebra $\Uq$ in the
module $V$. In particular, they provide explicit formulae for the action of
generators $t_m$, $e_m$ and $f_m$ on the vectors $\xiLa$ with $\La\in\TC$,
parallel to the formulae (\[3.z]),\,\(\[3.0]):
\vadjust{\kern-12pt}
$$
\gather
t_m\cdot\xiLa=\prod_{i=1}^mq^{\la_{mi}}\prod_{i=1}^{m-1}q^{-\la_{m-1,i}}
\,\xiLa\,,\\
\noalign{\vskip1pt}
e_m\cdot\xiLa=\sum_\Lap\,\ga_{mi\La}\,\tau_{mi\La}\,\xi_\Lap\,,\qquad
f_m\cdot\xiLa=\sum_\Lam\,\be_{mi\La}\,\tau_{mi\La}\,\xi_\Lam
\endgather
$$
where $\Lap$ and $\Lam$ are Gelfand-Zetlin schemes obtained from $\La$
by increasing and decreasing the $(m,i)$-entry by $1$ respectively.
The last two formulae look exactly as (\[3.0]).
\par
To obtain the $q$-analogues of the canonical Gelfand-Zetlin formulae given
in [J2] for the first time and rederived in [UTS], one should extend
the basic field $\F$ and to rescale the
vectors $\xi_\La$ in a way similar to that at the end of Section 3.
Namely, one should multiply $\xiLa$ by the factor
\vadjust{\kern1pt}
$$
\prod_{(l,j)}\kern6pt\Bigg\{\,\prod_{k=1}^{j-1}\,[\nu_{\,lj}-\nu_{\,lk}]
\cdot\kern-2pt
\prod_{s=\nu_j}^{\nu_{lj}-1}\,\Bigg(
\prod_{k=1}^j\,[s+1-\nu_k]\,\prod_{k=1}^{j-1}[s-\nu_k]\,
\prod_{k=j+1}^{l+1}\,[\nu_{\,l+1,k}-s-1]\,
\prod_{k=j}^{l-1}\,[\nu_{\,l-1,k}-s]\Bigg)\Bigg\}^{-1/2}\kern-17pt.
$$
\Demo{Remark 4.8}
Introduce the matrices
$$
Q=[q^{-2i}\de_{ij}]\ts_{i,j=1}^n\,,\qquad
\Th(x)=[\ts\Th_{ij}(x)\ts]\ts_{i,j=1}^n
$$
where $\Th_{ij}(x)$ is equal to $(-q)^{\,j-i}$ times the quantum determinant of
the matrix obtained from $T(x)$ by removing the row $j$ and the column $i$.
Then
$$
T(x)\ts\Th(x\kern.2pt q^{-2})=
Q\,\Th^{\,t}(x)\,Q^{-1}\,T^{\,t}(x\kern.2pt q^{2-2n})=\qd T(x)
\Tag{Last}
$$
where the superscript $t$ denotes the usual matrix transposition;
see [C1],[T] for the proof of these equalities.
Let $E_{\pm}$ be the matrices taking part in the homomorphism (\[Hom]):
$$
T(x)\mapsto x\kern.3pt E_{+}-E_{-}\,,
$$
cf.~the matrices $L^{(\pm)}$ from [RTF].
The coefficients of the polynomial\allowbreak
$a_n(x)=\qd(x\kern.3ptE_{+}-E_{-})$ belong to the center of the
algebra $\Uq$ and from (\[Last]) we obtain the matrix identities
$$
a_n(E_{-}E_{+}^{-1})=0\,,\qquad
a_n((E_{+}^{\,t})^{-1}E_{-}^{\,t\,}q^{2n-2})=0
$$
which are the $q$-analogues of (\[2.z]).
\enddemo

\Demo{Remark 4.9}
We can also treat $q$ as a complex number rather than an indeterminate and
consider $\Uq$, $\Yq$ as algebras over $\C$. If $q$ is generic then the
results of this section remain valid. In the peculiar case of $q$ being
a root of unit it is easy to generalize these results to the irreducible
highest weight modules. Moreover the technique works for the periodic and
semiperiodic modules [T] as well. This allows us to determine the
branching rules corresponding to the restriction from $\Uq$ to
$\U_q(\gl_{n-1})$ and to define the Gelfand-Zetlin bases for $\Uq$-modules
without classical analogues.
It will be done in details in the forthcoming paper.
\enddemo

\subheading{Acknowledgements}
We are grateful to Stanislav Pakuliak for discussions.
We should like to thank Masaki Kashiwara and Tetsuji Miwa for the hospitality
at RIMS.

\vskip\baselineskip

\parindent13.3pt
\line{\kern2\parindent\kern-.5pt\bf References \hfil}\medskip\smallskip

\itemitem{[C1]}
I. V. Cherednik,
{\sl A new interpretation of Gelfand-Zetlin bases},
Duke Math. J.
{\bf 54}
(1987)
563--577.

\itemitem{[C2]}
I. V. Cherednik,
{\sl Quantum groups as hidden symmetries of classic representation theory},
in
`Differential geometric methods in theoretical physics (A. I. Solomon, Ed.)',
World Scientific,
Singapore,
1989,
pp. 47--54.

\itemitem{[D1]}
V. G. Drinfeld,
{\sl Hopf algebras and the quantum Yang--Baxter equation},
Soviet Math. Dokl.
{\bf 32}
(1985)
254--258.

\itemitem{[D2]}
V. G. Drinfeld,
{\sl A new realization of Yangians and quantized affine algebras},
Soviet Math. Dokl.
{\bf 36}
(1988)
212--216.

\itemitem{[G]}
M. D. Gould,
{\sl On the matrix elements of the $U(n)$ generators},
J. Math. Phys.
{\bf 22}
(1981)
15--22.

\itemitem{[GZ]}
I. M. Gelfand, M. L. Zetlin,
{\sl Finite-dimensional representations of the unimodular group},
Dokl. Akad. Nauk SSSR
{\bf 71}
(1950)
825--828.

\itemitem{[J1]}
M. Jimbo,
{\sl A $q$-analogue of $U(gl(N+1))$, Hecke algebra and the Yang-Baxter
equation},
Lett. Math. Phys. {\bf11} (1986) 247--252.

\itemitem{[J2]}
M. Jimbo,
{\sl Quantum $R$-matrix for the generalized Toda system},
Commun. Math. Phys. {\bf102} (1986) 537--547.

\itemitem{[KS]}
P. P. Kulish, E. K. Sklyanin,
{\sl Quantum spectral transform method: recent developments},
in `Integrable quantum field theories',
Lecture Notes in Phys.
{\bf 151},
Springer,
Berlin--Heidelberg,
1982,
pp. 61--119.

\itemitem{[L]}
G. Lusztig,
{\sl On deformations of certain simple modules over enveloping algebras},
Adv. Math.
{\bf70}
(1988)
237--249.

\itemitem{[N]}
M. L. Nazarov,
{\sl Quantum Berezinian and the classical Capelli identity},
Lett. Math. Phys.
{\bf 21}
(1991)
123--131.

\itemitem{[O]}
G. I. Olshanski,
{\sl Representations of infinite-dimensional classical groups,
limits of enveloping algebras, and Yangians},
in `Topics in Representation Theory (A. A. Kirillov, Ed.)',
Adv. Soviet Math.
{\bf 2},
AMS,
Providence RI,
1991,
pp. 1--66.

\itemitem{[R]}
M. Rosso,
{\sl Finite-demensional representations of the $q$-analogues of the
universal enveloping algebras of complex simple Lie algebras},
Commun. Math. Phys.
{\bf117}
(1988)
581--593.

\itemitem{[RTF]}
N.Yu. Reshetikhin, L.A. Takhtajan and L.D. Faddeev,
{\sl Quantization of Lie groups and Lie algebras},
Leningrad Math. J.
{\bf1}:1
(1990)
193--225.

\itemitem{[T]}
V. Tarasov,
{\sl Cyclic monodromy matrices for $sl(n)$ trigonometric $R$-matrices},
Preprint RIMS
{\bf 903}
(1992)
1--34.

\itemitem{[UTS]}
K. Ueno, T. Takebayashi and Y. Shibukawa,
{\sl Gelfand-Zetlin basis for $U_q(gl(N+1))$-modules},
Lett. Math. Phys.
{\bf18}
(1989)
215--221.

\itemitem{[Z1]}
D. P. Zhelobenko,
{\sl Compact Lie groups and their representations},
Transl. of Math. Monographs
{\bf 40},
AMS,
Providence RI,
1973.

\itemitem{[Z2]}
D. P. Zhelobenko,
{\sl On Gelfand-Zetlin bases for classical Lie algebras},
in `Representations of Lie groups and Lie algebras (A. A. Kirillov, Ed.)',
Akad\'emiai Kiad\'o, Budapest, 1985, pp. 79--106.

\enddocument
\bye